\documentclass[pre,twocolumn,showpacs,aps,floats,floatfix,superscriptaddress]{revtex4}
\usepackage{epsfig}
\usepackage{subfigure}

\newcommand{\bm}[1]{\mbox{\boldmath$#1$}}

\begin{document}
\title{Confined nanorods: jamming due to helical buckling}
\date{\today}

\author{Daniel Sven\v sek}

\address{Department of Physics, Faculty of Mathematics and Physics, University of Ljubljana, Jadranska 19, SI-1111 Ljubljana, Slovenia}

\author{Rudolf Podgornik}
\address{Department of Physics, Faculty of Mathematics and Physics, University of Ljubljana, Jadranska 19, SI-1111 Ljubljana, Slovenia}
\address{Department of Theoretical Physics, J. Stefan Institute, Jamova 39, SI-1000 Ljubljana, Slovenia}

\begin{abstract}
We investigate a longitudinally loaded elastic nanorod inside a cylindrical channel and show within the context of classical elasticity theory that the Euler buckling instability leads to a helical postbuckling form of the rod within the channel. The local pitch of the confined helix changes along the channel and so does the longitudinal force transmitted along the rod, diminishing away from the loaded end. This creates a possibility of jamming of the nanorod within the channel.
\end{abstract}


\maketitle

\section{Introduction}

Elastic instabilities of nanoscale rods and in particular the Euler (buckling) instability have long been recognized as being essential for structural as well as functional aspects of nano- and bio-systems. 
Some time ago  \cite{manning} Manning proposed that DNA collapse in polyvalent salt solutions proceeds through a type of Euler buckling driven by diminished electrostatic repulsive interactions between charged phosphates along the DNA backbone. Though the details of his picture appear to be speculative, the general framework has been completely vindicated by later work \cite{podgornik-1}. Buckling instability appears to play also an important part in the conformation of linker DNA within the chromatin fiber \cite{lesne}, in the case of a growing microtubule pushing  against a microfabricated rigid barrier \cite{dogterom-1} and in the stability of AFM-tips, nanotubes \cite{nanotube-inst}  and nanorods \cite{nanorod-inst}. 

The confinement of nanoscale rods to micro- and nano-channels presents yet another set of experimental and theoretical problems \cite{dna-nanochannel} centered around the technological challenges 
of manufacturing nanopores and nanochannels for investigating and manipulating DNA. Here the main theoretical thrust is in the direction of understanding the various contributions to the confinement free energy that depend on the size of the confining space and the intrinsic properties of the confined nanorods. The behavior of confined semiflexible polymers is particularly important in this context and has been recently analyzed in detail \cite{nanochan,odijk}. In many respects this type of problems are mainly centered upon transverse confinement, whereas elastic instabilities described before are due to longitudinal confinement of elastic rods.

Motivated by these phenomena we will try in what follows  to combine the two aspects of polymer confinement described above, considering an elastic buckling instability within a confining cylindrical channel. The confinement of the elastic rod is thus twofold: longitudinal, leading to buckling, and transverse, leading straightforwardly to a helical postbuckling form. We will show that transverse confinement with frictional walls furthermore leads to a decay of the longitudinal force within the rod along the long axis of the confining channel.
This will not be too difficult to rationalize since we know from other systems that longitudinal stresses can be taken up effectively by friction at confining walls. For a silo filled with granular matter and compressed on one side by a force $F_0$, Janssen's equation \cite{janssen} gives an exponential decrease of the compression force with the distance $z$ from the compressed side, $$F(z) = F_0\, {\rm exp}(-\lambda k_{fr} z {\cal P}/{\cal S}),$$where $\lambda = \sigma_{rr}/\sigma_{zz}$ is the ratio of horizontal and vertical stresses, ${\cal P}/{\cal S}$ is the perimeter to cross-section area ratio, $k_{fr}$ is the static friction coefficient and we have neglected gravity for the purpose. The same holds also for a regular elastic solid, i.e., a rigidly confined rod, if one neglects lateral deformation gradients in the rod \cite{elastic_validity}. 
There we have $$F(z) = F_0\, {\rm exp}[-\sigma k_{fr} z {\cal P}/{\cal S}(1-\sigma)],$$where $\sigma$ is Poisson's ratio. On this simple basis one expects also a conceptually similar effect, i.e., a decrease of compression force with the distance along an elastic rod confined to a cylindrical channel  in the post-buckling regime. This will allow us to hypothesize on the existence of a jamming regime for sufficiently long confined elastic rods.

The plan of this paper is as follows: we shall first present the classical elastic model for an Eulerian rod and derive the scaling relations valid for the onset of the buckling instability. We shall solve an approximate form of the elastic equations within a cylindrically confined channel and show that the longitudinal force, at least within the considered approximations, decays inversely proportional to the length of the confining channel when friction at the cylindrical wall is taken into account. To obtain exact solutions of the elastic equations we will proceed numerically and derive the minimal shape of the confined rod and the magnitude of the transmitted longitudinal force along the axis of the confining channel. At the end we will discuss what could be the conditions that would lead to a jamming of the elastic rod within the cylindrical confining channel.

\section{Elastic model}

\noindent
We model the rod as a thin Eulerian elastic filament to which the standard continuum theory of elasticity can be applied. The confining channel  is modeled as a straight cylindrical tube with rigid walls. The ratio $R/L$ of tube radius $R$ and undeformed filament length $L$ is the relevant geometric parameter of the system.

Let us briefly review the equilibrium equations for thin elastic rods \cite{landau}.
The force balance reads
\begin{equation}
	{{\rm d}{\bf F}\over{\rm d}l} + {\bf K} = 0,
	\label{force-bal}
\end{equation}
where ${\rm d}l$ is the length element on the rod ($l$ is the natural parameter---the length along the filament), ${\bf F}$ is the elastic force exerted on the leading surface of the rod element (in the sense of increasing $l$; the force on the opposite (trailing) side is $-{\bf F}$, of course), and ${\bf K}$ is the external force per unit length.
The torque balance reads
\begin{equation}
	{{\rm d}{\bf M}\over{\rm d}l} + {\bf t}\times{\bf F} = 0,
	\label{torque-bal}
\end{equation}
where $\bf M$ is the elastic torque exerted about the center of the leading surface and $\bf t$ is the unit tangent of the rod. 
The elastic torque is related to the deformation of the filament via
\begin{equation}
	{\bf M} = G \tau\, {\bf t} + E I\, {\bf t}\times{\dot{\bf t}},
	\label{M_complete}
\end{equation}
where $E$ is the Young modulus and we assumed that the rod is circular, i.e., both geometric moments of inertia of the cross section are equal, $I = \int\! {\rm d}S\, x^2  = \int\! {\rm d}S\, y^2$.
The first term on the right describes torsion which we assume to be absent. One can show that in case of the isotropic moment of inertia, torsion is absent everywhere in the rod when there is no twisting torque applied \cite{landau}. In what follows, the twisting (torsional) torque will be absent, i.e., the torque in the rod will have a vanishing tangential component,
\begin{equation}
	{\bf M} = E I\, {\bf t}\times{\dot{\bf t}}.
	\label{M}	
\end{equation}
From Eqs.~(\ref{torque-bal})~and~(\ref{M}) one then gets
\begin{equation}
	EI\, {\bf t}\times\ddot{\bf t} + {\bf t}\times{\bf F} = 0
	\label{dM/dl}
\end{equation}
as the fundamental equation describing the shape of the bent filament. The solutions of this equation with various boundary conditions are discussed in standard references on the theory of elasticity \cite{landau}.

\subsection{Buckling in confined geometry}

\noindent
It is well known that when a straight rod is compressed by an axial force, above threshold it undergoes a buckling instability---the so-called Euler buckling---and becomes bent. For a rod with hinged ends or ends that are laterally free (both situations are identical in this case), for example, the threshold force is $F_{crit}= E I\pi^2/L^2$ \cite{landau}. 

What happens when the (thin) rod is confined within a rigid cylinder? Once the compressing force is sufficiently large so that the bent rod touches the wall of the cylinder (let it be bent in the diameter plane) and is further increased, there are two scenarios one can think of: the rod can bend back-and-forth staying in the diameter plane, or it can become a helix touching the cylinder wall (spring-like configuration). In oil-drilling community it has been recognized for a long time  that it is the latter case that eventually happens \cite{lubinski,mitchell1988,mitchell1996,mitchell2002} (In fact, this is the only area of research to have studied helical buckling.) It turns out, as we will show in Section~\ref{results}, that there indeed exists a secondary threshold above which the planar configuration is unstable with respect to the helix.

\subsection{Scaling}

\noindent
By introducing dimensionless quantites denoted with tilde,
\begin{equation}
	 l = R\, \tilde{l},\quad 
	 {\bf F} = {E I\over R^2}\,{\bf\tilde{F}},\quad
	 {\bf K} = {E I\over R^3}\,{\bf\tilde{K}},\quad	 
	 {\bf M} = {E I\over R}\,{\bf\tilde{M}},
	 \label{scaling}
\end{equation}
Eqs.~(\ref{force-bal})-(\ref{dM/dl}) attain a universal form. Thus, the solutions depend only on the aspect ratio $L/R$. Upon rescaling the system size and preserving the aspect ratio, the solution is unchanged if also the forces and torques of the boundary condition are rescaled correspondingly, i.e., according to Eq.~(\ref{scaling}). Similarly, taking a rod with a different $E I$ and rescaling the forces and torques does not affect the solution.

\section{Analytic description of the helix}

\subsection{Simple model}
\label{simple_model}

\noindent
Let us write down equilibrium equations for a helically deformed filament constrained to and lying on a cylindrical surface  of radius $R$. The deformation is sustained by an external compressing force $F_z<0$ parallel to the long  axis of the cylinder. We disregard the transition regions in the vicinity of both ends and focus on the central part of the long helix, i.e., we assume that its pitch is constant. Friction is absent at this stage.

In cylindrical coordinates $(r,\phi,z)$
the helix is given by $r=R$ and the linear function $\phi(l)$. The tangent is thus given by
\begin{equation}
	{\bf t} = {{\rm d}{\bf r}\over{\rm d}l} = R \dot{\phi}\,\hat{\bf e}_\phi + \sqrt{1-(R \dot{\phi})^2}\, 
		  \hat{\bf e}_z. 
	\label{tangent}
\end{equation}
Further we put
${\bf F} = F_r \hat{\bf e}_r+F_\phi \hat{\bf e}_\phi+F_z \hat{\bf e}_z$, where the first two components are unknown and must be determined. The force of the cylinder per unit length is
${\bf K} = -K \hat{\bf e}_r$.
Eq.~(\ref{dM/dl}) then reads
\begin{widetext}
\begin{equation}
	\left[\sqrt{1-(R \dot{\phi})^2}(EI\,R\dot\phi^3-F_\phi)+R\dot\phi F_z \right]\hat{\bf e}_r + 
	\sqrt{1-(R \dot{\phi})^2}F_r\, \hat{\bf e}_\phi - R\dot\phi F_r\, \hat{\bf e}_z = 0.
	\label{equi-components}
\end{equation}
\end{widetext}
Eq.~(\ref{equi-components}) requires 
\begin{equation}
	F_r=0,
	\label{Fr-equi}
\end{equation} 
while from Eq.~(\ref{force-bal}) one gets 
\begin{equation}
	F_\phi = - {K\over\dot\phi}
	\label{Fphi-equi}
\end{equation}
and hence for the radial component of Eq.~(\ref{equi-components})
\begin{equation}
	\sqrt{1-(R \dot{\phi})^2}\left(EI\, R\dot\phi^4 +  K\right) + R\dot\phi^2 F_z = 0.
	\label{equilibrium}
\end{equation}

Now comes a crucial point.
The equilibrium condition (\ref{equilibrium}) involves both the helical deformation $\dot\phi$ and 
the external force $\bf K$, so additional input is needed to determine one or the other.
This comes from the boundary condition, which is however inaccessible under the assumption of fixed pitch. 

Let us digress a little and illustrate in physical terms why the force of the cylinder, $\bf K$, cannot be specified until the boundary condition is known. Besides the force $\bf F$, the boundary condition involves also the torque exerted on free ends (which must be normal to the tangent of the rod as we are not considering torsion). By using a pair of wrenches and applying torque to the ends, one changes the force exerted on the cylinder wall. Using just the right torque, for example,
one can make $\bf K$ vanish. Increasing the torque further, the rod detaches from the cylinder and forms a free helix with the radius smaller than $R$ \cite{love}.
According to Eqs.~(\ref{M})~and~(\ref{tangent}), the torque in the filament is 
\begin{equation}
	{\bf M} = EI\,R\dot\phi^2\left[-\sqrt{1-(R\dot\phi)^2}\,\hat{\bf e}_\phi + R\dot\phi\,\hat{\bf e}_z\right].
	\label{M-helix}
\end{equation}
In fact, with this torque applied to the ends (together with $F_\phi$ and $F_z$), the helical pitch is constant everywhere. By changing the torque, one changes the pitch, Eq.~(\ref{M-helix}), and therewith the cylinder force $\bf K$, Eq.~(\ref{equilibrium}). Note that there is no tangential force in the free (detached from the cylinder) helix, Eq.~(\ref{Fphi-equi}). 

\begin{figure}[h]
\begin{center}
	\mbox{\subfigure[]{\includegraphics[width=4.2cm]{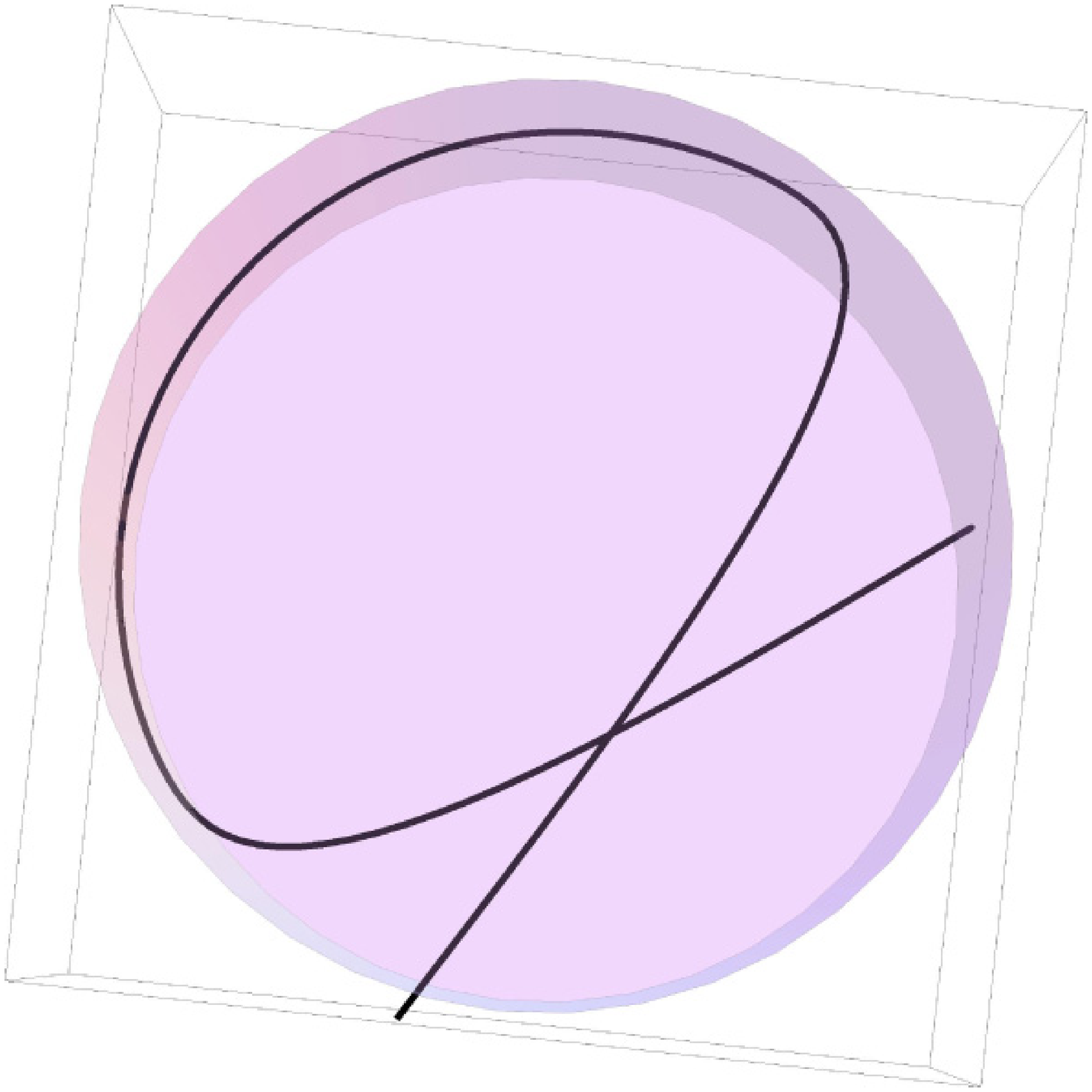}}\hspace{0cm}
		  \subfigure[]{\includegraphics[width=4.2cm]{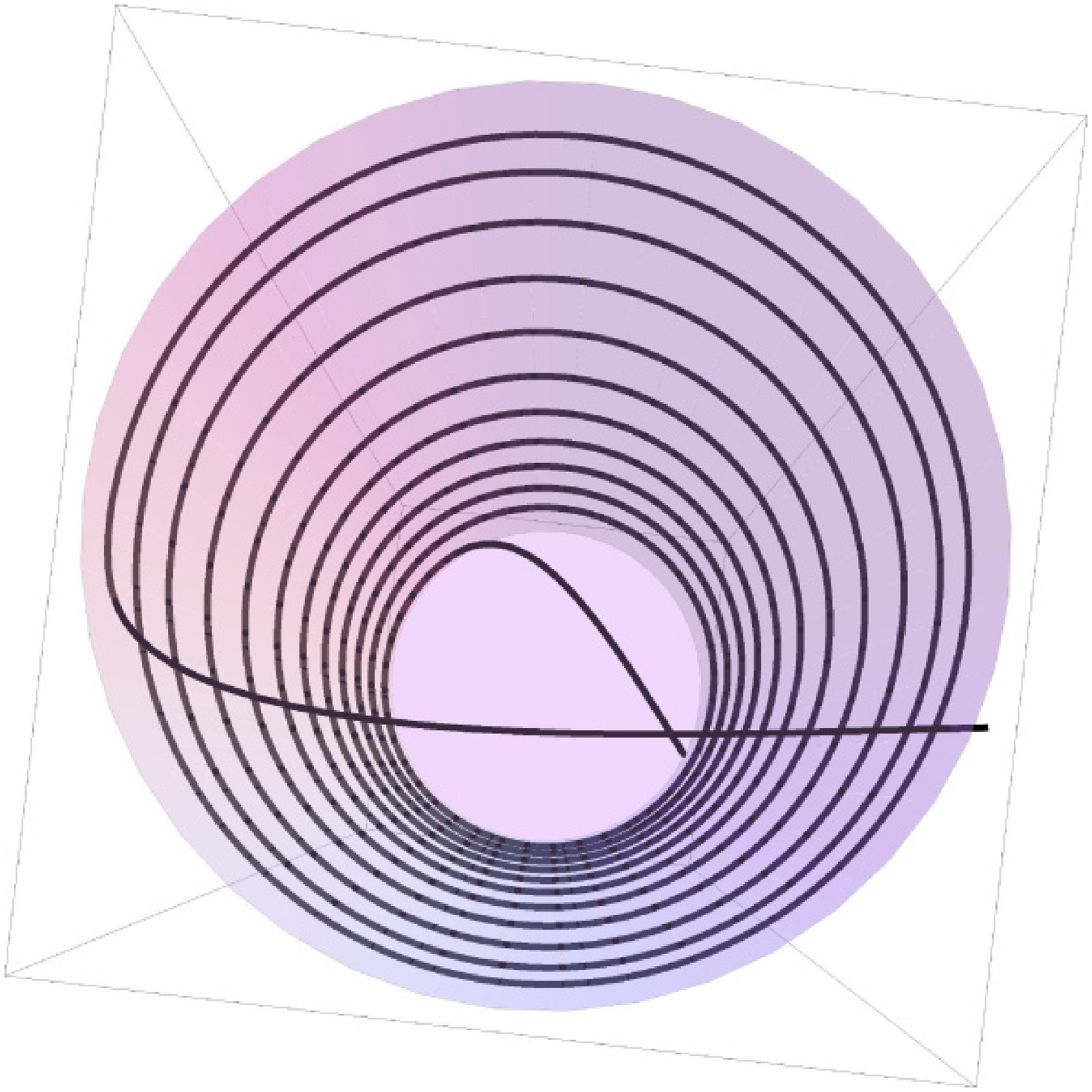}}\hspace{0cm}  }
\caption{(color online) A long-axis view of helical filaments with lengths (a) $L/R = 20$ and (b) $L/R = 640$, buckled by a force $\vert F_z(0)\vert=0.25 {E I/R^2}$, which is approximately (a) 10-times and (b) 10,000-times the critical force for Euler buckling $F_{crit}= E I\pi^2/L^2$. There is no torque applied to the ends.
\label{short-long}}
\end{center}
\end{figure}

When there is no torque applied to the ends, i.e., when compressing the rod inside the cylinder by a pair of pistons, the part of the rod close to the end detaches from the wall while the tip (which is straight due to zero torque) pushes against the wall with a discrete force (see numeric solutions, Fig.~\ref{short-long}). The torque due to this force increases as we move away from the tip and eventually becomes sufficiently large for the helical deformation.

Let us return to the case of a constant helical pitch. One can invoke an energy argument to independently estimate the dependence of $\dot\phi$ on the compressing force $F_z$. 
The elastic free energy of the compressed helix without the contribution of the torsion is
\begin{equation}
	{\cal F} = {1\over 2}EI\int_0^{L}\!\!{\rm d}l\left\vert \dot{\bf t}\right\vert^2 - L_z F_z,
\end{equation}
where $L$ is the length of the filament and $L_z$ is the length of its projection to the $z$ axis, i.e., the length of the coil. Recall that $F_z<0$. With Eq.~(\ref{tangent}) and ${\rm d}z/{\rm d}l = \sqrt{1-(R\dot\phi)^2} = L_z/L$, we have
\begin{equation}
	{\cal F} = {1\over 2}EIL\, (R\dot\phi^2)^2 + L\sqrt{1-(R\dot\phi)^2}\, \vert F_z\vert
\end{equation}
and after minimizing, in the lowest order of $R\dot\phi$,
\begin{equation}
	\dot\phi^2 = {\vert F_z\vert\over2EI}.
	\label{dotphi}
\end{equation}
Hence, from Eq.~(\ref{equilibrium}) one gets a direct relation between the load and the wall forces. To the lowest order the force of the cylinder is
\begin{equation}
	K = {R\over 2EI}F_z^2.
	\label{K-simple}
\end{equation}

So far the friction has been absent. Let us now consider it as a perturbation and introduce a small friction coefficient $k_{fr}$. One is aware, of course, that for finite friction the constant pitch is not a solution, neither can one use the energy argument.
We are interested in how the compressing force $\vert F_z\vert$ is reduced by the friction going from one end of the helix to the other.
Let us assume that the helix is compressed by the external force $F_{z0}$ at $l=0$ and that the friction is parallel to $z$, opposing the external force.
Assuming the static friction is maximum everywhere, it follows from the force balance (\ref{force-bal}) that
\begin{equation}
	{{\rm d}\vert F_z\vert\over{\rm d}l} = -k_{fr} K.
	\label{Fz-simple}
\end{equation}
Combining Eqs.~(\ref{K-simple})~and~(\ref{Fz-simple}) we finally get
\begin{equation}
	\vert F_z(l)\vert = \left({1\over \vert F_{z0}\vert}+{k_{fr}  R\over 2EI}\,l\right)^{-1}.
	\label{Fz-simple}
\end{equation}
The virtual point where $F_z$ would diverge lies at 
\begin{equation}
	l_c=-2EI/k_{fr} R\vert F_{z0}\vert, 
	\label{l_c}
\end{equation}
which thus dictates the extent of the reduction of the compression force. The smaller $\vert l_c\vert$, the strongest is the decay of the force.
Increasing $k_{fr}$, $\vert F_{z0}\vert$, or $R$ reduces $\vert l_c\vert$ and moves the singular point closer to the beginning of the helix.
Eq.~(\ref{Fz-simple}) can be put into a universal form by introducing the force unit $F_0 = 2EI/k_{fr}RL$,
\begin{equation}
	{\vert F_z(l)\vert\over F_0} = \left({F_0\over \vert F_{z0}\vert}+{l\over L}\right)^{-1}.
	\label{Fz-universal}
\end{equation}
Note in Eq.~(\ref{scaling}) that if the aspect ratio is constant, the force scales as $1/R^2$ (or equivalently, $1/L^2$). This, of course, always holds and is approximation-independent. In addition, the result (\ref{Fz-universal}) of our simple model gives a $1/RL$ scaling of the force when the aspect ratio is changed, which, however, is only an approximation.

\subsection{Exact analysis}

\noindent
For completness, let us derive exact equilibrium equations for the helically buckled filament in the presence of friction. 
We do not aim to solve them, yet they will be helpful giving us insight into the solution.
As before, the filament is assumed to lie on the cylinder everywhere, $r=R$, but now all variables are $l$-dependent, including the compressing force $F_z$.

Eq.~(\ref{dM/dl}) involves only two components as it is orthogonal to $\bf t$. Therefore we project it to $\hat{\bf e}_r$ and the direction $\bf t \times\hat{\bf e}_r$. Writing $\dot\phi(l)=\omega(l)$,
the former is
\begin{widetext}
\begin{equation}
	{EIR\over\left(1-R^2\omega^2\right)^{3/2}}\left[\omega^3\left(1-R^2\omega^2\right)^2-
	\ddot\omega\left(1-R^2\omega^2\right)-R^2\omega\dot\omega^2\right]-F_\phi\sqrt{1-R^2\omega^2}+F_z R\omega=0
	\label{M-er}
\end{equation}
\end{widetext}
and the latter
\begin{equation}
	-{3\over 2}EIR\,{{\rm d}\over{\rm d}l}(\omega^2) + F_r = 0.
	\label{M-e2}
\end{equation}

Again we choose from an infinite number of solutions, which exist when the static friction is involved, the simplest and most symmetric one, in which the friction is everywhere at its maximum and parallel to $z$.
With ${\bf K}=K(l)(-\hat{\bf e}_r + k_{fr}\hat{\bf e}_z)$ the three components of the force balance (\ref{force-bal}) are
\begin{eqnarray}
	\left(\dot{F}_r-\omega F_\phi -K\right)\hat{\bf e}_r &=& 0,\label{F-er}\\
	\left(\omega F_r+\dot{F}_\phi\right)\hat{\bf e}_\phi &=& 0,\label{F-ephi}\\
	\left(\dot{F}_z + k_{fr}  K\right)\hat{\bf e}_z &=& 0. \label{F-ez} 
\end{eqnarray}
Eqs.~(\ref{M-er})-(\ref{F-ez}) represent a closed set of five ordinary differential equations for the five variables $\omega(l)$, $K(l)$, and ${\bf F}(l)$. One can verify that in case $\dot\omega=0$, Eqs.~(\ref{Fr-equi})-(\ref{equilibrium}) are recovered.

One can expand the system (\ref{M-er})-(\ref{F-ez}) for small $R\omega$. For $k_{fr}\ne 0$, terms of third order must be included to get a solution (one can check that the second order gives only the second order part of the stationary solution (\ref{Fr-equi})-(\ref{equilibrium}) for $k_{fr}=0$, and no solution for $k_{fr}\ne0$).
Inserting Eq.~(\ref{M-e2}) into Eq.~(\ref{F-ephi}) and integrating, one gets a partial result $F_\phi = A - EI\omega^3$,
where $A$ is a constant.

\section{Numerical approach}

\noindent

We will solve the fundamental equations for the actual deformation of the compressed and confined filament numerically for it is only in this way that one can obtain a complete solution of the problem. The filament is naturally allowed to detach from the wall, which essentially takes place near the two ends of the filament, and the helix within the cylindrical pore can form spontaneously without having been put in "by hand". Let us stress again that the solution, including in particular the pitch of the bulk helix and the force exerted on the cylinder wall, is only determined when the ends of the filament  are taken into account, which furthermore inevitably involve detachment of the filament from the wall unless the boundary condition is very special, as explained in Sec.~\ref{simple_model}, Eq.~(\ref{M-helix}).

For the purpose of numeric modelling that involves discretization, it is generally better, if only possible, to start with a discrete analogue of the continuum system and write down algebraic equations, rather than discretizing the differential equations themselves, derived for the continuum limit. In this spirit, the elastic filament will be represented as a set of straight and stiff elements (links) of fixed length $l_0$ with forces and torques acting between them. The element $i$ is described by its center of mass ${\bf r}_i$ and a unit (tangent) vector ${\bf t}_i$ giving its orientation. Dynamical evolution of the filament shape is obtained by Newton's laws for translation and rotation of each element.
\begin{figure}[h]
 \begin{center}
 \includegraphics[width=9cm]{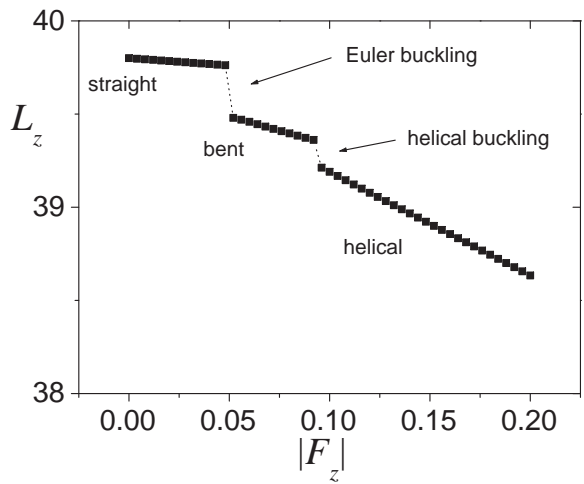}
 \caption{The external load dependence of the projection of filament length onto the axis of the confining cylindrical pore. The weak force dependence of length in the straight configuration is due to the small artificial compressibility introduced by the penalty potential (\ref{penalty}) (i.e., finite value of the spring constant $k$). In the limit of zero compressibility this part of the functional dependence would be a straight horizontal line.
 \label{bifurcation-diagram}}
\end{center}
\end{figure}
The joints between the links are rather cumbersome to model: they act as constraints for the ends of the two elements which must meet in a single point, and thus exert constraint forces that can be determined only implicitly, such that the elements satisfy the constraints. To avoid this, we relax (soften) the constraints and introduce a quadratic (bond) penalty potential 
\begin{equation}
	V({\bf r}_i^+,{\bf r}_{i+1}^-) = {k\over 2}\left\vert{{\bf r}_{i+1}^- -{\bf r}_i^+}\right\vert^2,
	\label{penalty}
\end{equation}
where ${\bf r}_i^+ = {\bf r}_i+(l_0/2){\bf t}_i$ and ${\bf r}_{i+1}^- = {\bf r}_{i+1}-(l_0/2){\bf t}_{i+1}$ are the two ends meeting at the joint. The coefficient $k$ should be sufficiently large so that the gap between the ends is small compared to the element length, $\left\vert{{\bf r}_{i+1}^- -{\bf r}_i^+}\right\vert \ll l_0$. The side wall of the cylinder is modelled in a similar manner, introducing a wall penalty potential
\begin{equation}
	V_w({\bf r}_i^{\pm}) = \left\{
	\begin{array}{lcc}
	{k_w\over 2}\left(\vert{\bf r}_i^{\pm}\vert-R\right)^2 &; & \vert{\bf r}_i^{\pm}\vert>R\\
	0 &; & \vert{\bf r}_i^{\pm}\vert\le R
	\end{array}
	\right..
\end{equation}
Again the coefficient $k_w$ should be sufficiently large so that $\vert{\bf r}_i^{\pm}\vert-R\ll l_0$. 
The forces on the ends of the elements are 
\begin{equation}
	{\bf F}_i^\pm = -{\partial V\over\partial{\bf r}_i^\pm} - {\partial V_w\over\partial{\bf r}_i^\pm}.	
\end{equation}
It makes little difference (i.e., no difference in the limit $l_0\to 0$) whether the force of the wall actually acts on both ends or just at the center of the element.

Torques about the center of element $i$ are of two kinds. One comes from elastic couples exerted by the two neighbouring elements and the other from the forces on both ends of element $i$:
\begin{equation}
	{\bf M}_i = C\, {\bf t}_i\times\left({\bf t}_{i-1}+{\bf t}_{i+1}\right) + 
		    {l_0\over 2}\left({\bf t}_i\times{\bf F}_i^+ - {\bf t}_i\times{\bf F}_i^-\right),
	\label{M-discr}
\end{equation}
where $C$ is the bending stiffness which we shall connect to continuum parameters. In the continuum picture, the first part of the torque translates to the first term of Eq.~(\ref{dM/dl}), and the second part to the second term of Eq.~(\ref{dM/dl}).

To get the evolution of the shape of the filament, we coveniently assume overdamped dynamics:
\begin{eqnarray}
	\beta {{\rm d}{\bf r}_i\over{\rm d}t} &=& {\bf F}_i^+ + {\bf F}_i^-,\label{transl}\\
	\beta {{\rm d}{\bf t}_i\over{\rm d}t} &=& \beta {{\rm d}{\bm \varphi}_i\over{\rm d}t}\times{\bf t}_i = 
		{\bf M}_i\times{\bf t}_i,\label{rot} 
\end{eqnarray}
where $\beta$ is an arbitrary damping coefficient defining the time scale.

One can verify that the set of discrete equations (\ref{M-discr})-(\ref{rot}) agrees with properly discretized continuum equations with the following connection between the parameters:
\begin{equation}
	C = {EI\over l_0^2},
\end{equation}
while $\beta=l_0\beta_{ph}$, where $\beta_{ph}$ is the physical (continuum) damping coefficient. The latter is only important for properly scaling the time when changing the element length $l_0$ in the numerical model. Similarly, $k=k_{ph}/l_0$ to ensure that the spring constant $k_{ph}$ of the rod (which is infinite in the continuum description (\ref{force-bal})-(\ref{M}) of a thin rod) is unaffected by changes of $l_0$.

To avoid giant friction forces in certain discrete points (e.g., a few points arround the detachment region which exert a large force on the confining wall), the friction force is properly capped, so that the Coulomb's friction law is violated in these few points but met everywhere else.

\section{Results}
\label{results}

First let us take a look at numeric solutions of helically buckled filaments. 
Fig.~\ref{short-long} shows two examples of filaments of different lengths buckled by the same axial force. We use the most natural boundary condition: the upper (compressed) end is only pushed along $z$ (the cylinder axis), while the lower end is not allowed to move in $z$ direction. Both ends are free to move laterally until they hit the confining wall. There is no torque applied to them. To build up the torque exerted on the filament cross section in the helical state, however, the ends must leave the helical state as anticipated in Sec.~\ref{simple_model}.
\begin{figure}[h]
\begin{center}
	\parbox{10cm}{\hspace{-1.4cm}\subfigure[]{\includegraphics[width=4.3cm]{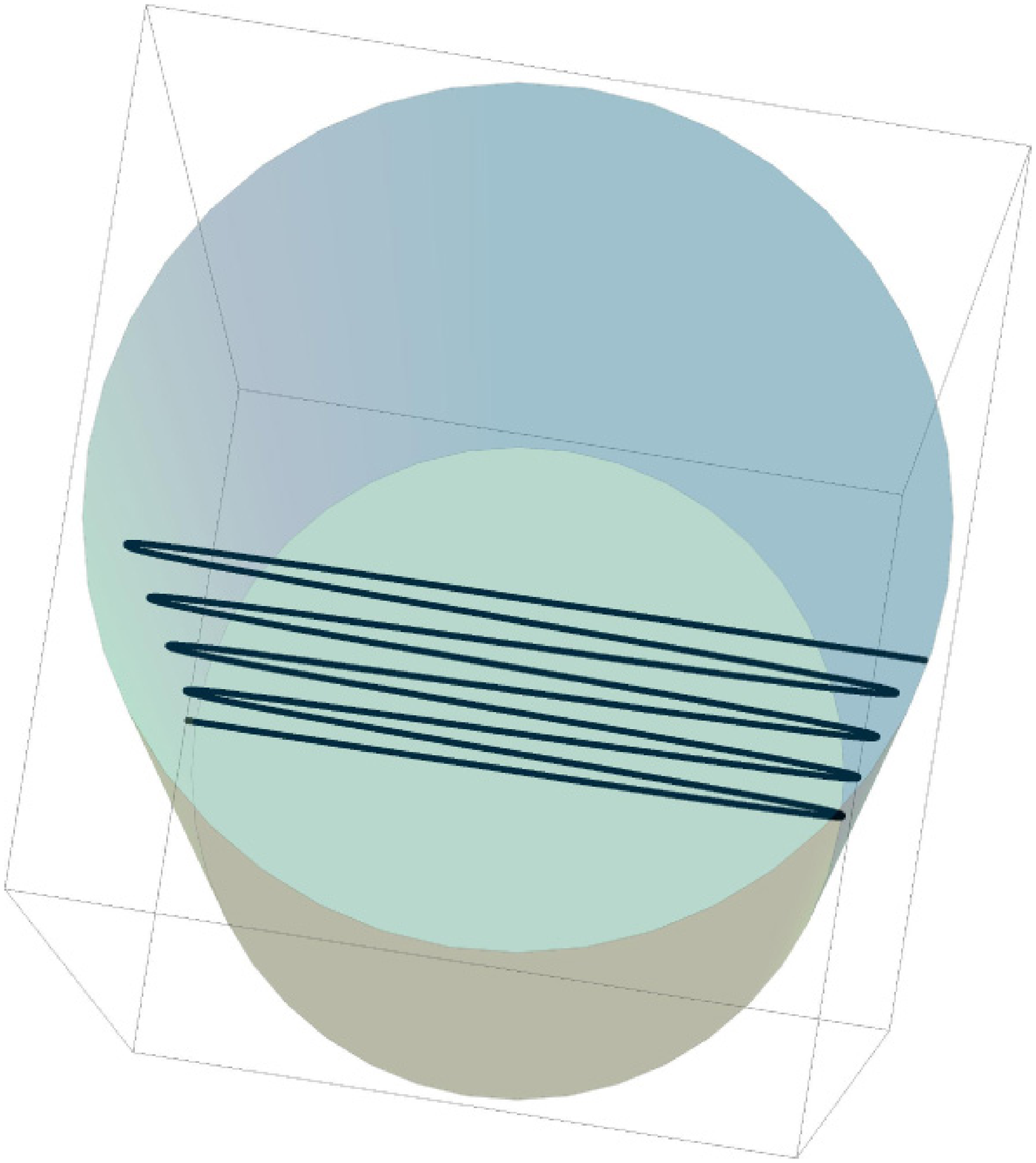}}\hspace{-0.3cm}
		  \subfigure[]{\includegraphics[width=4.3cm]{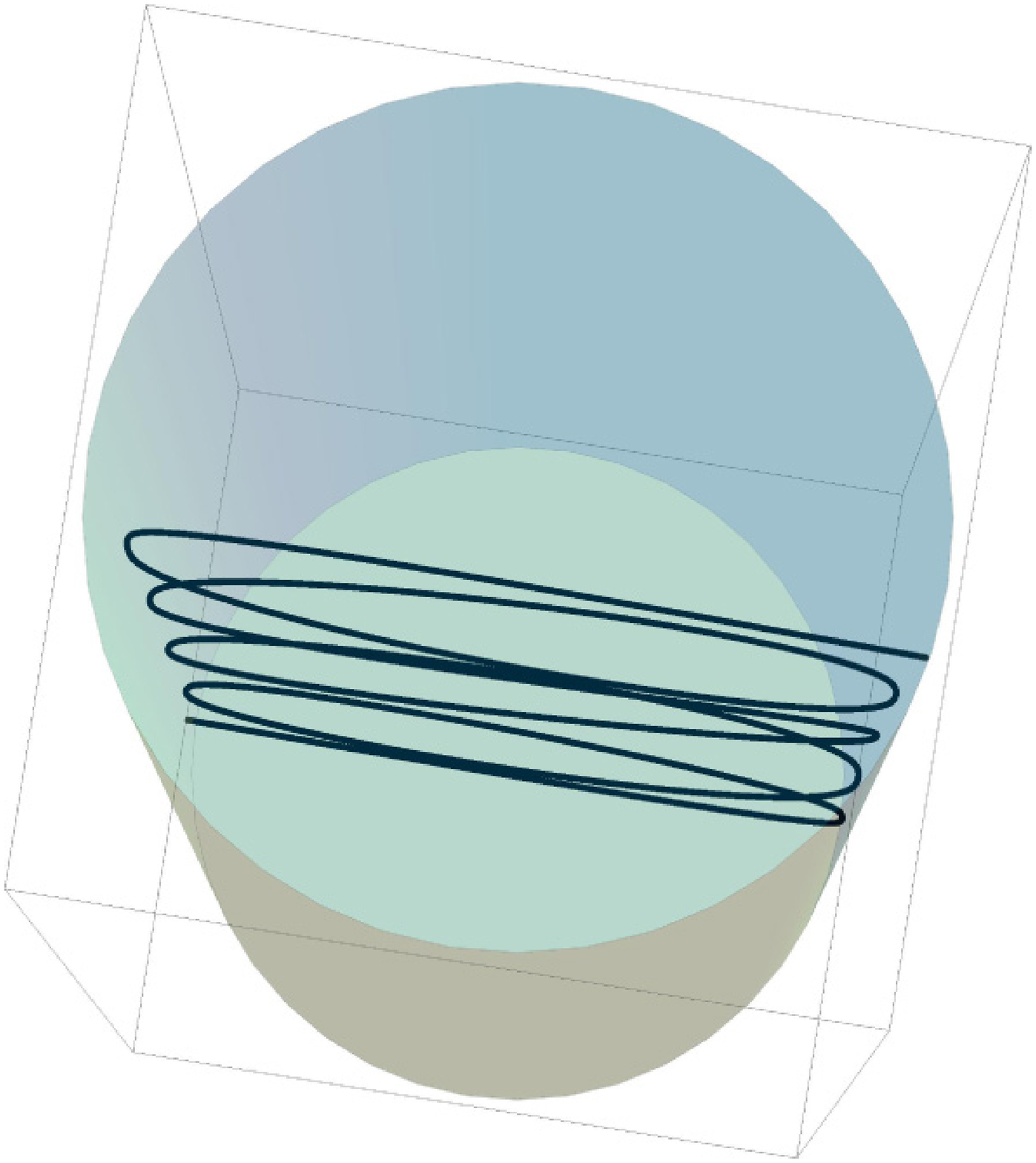}}\hspace{0cm}\\
		  \hspace{-1.4cm}\subfigure[]{\includegraphics[width=4.3cm]{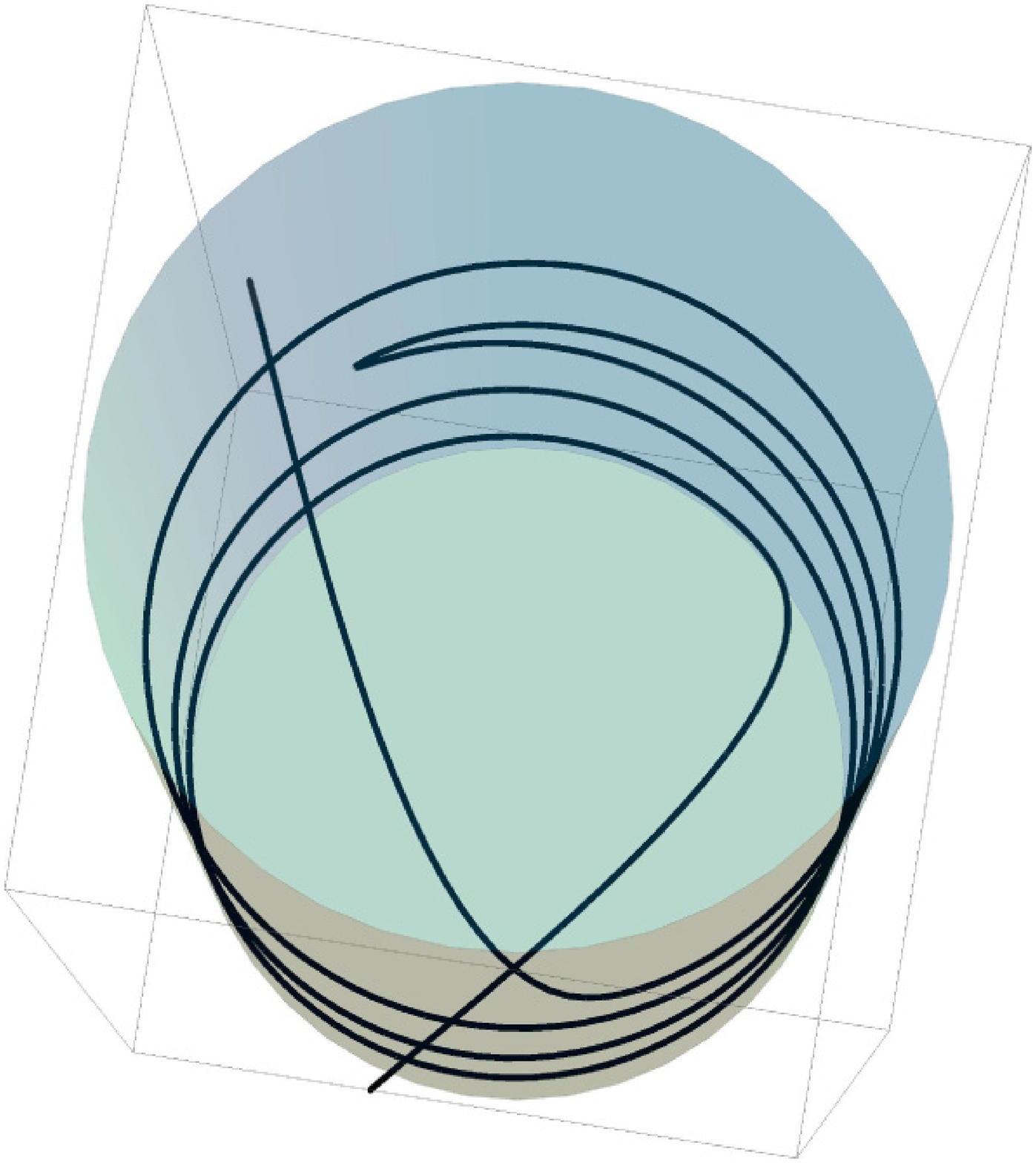}}
		  }
\caption{(color online) Planar configuration of the confined filament (a) is unstable (b) and is transformed into the helical (c) upon perturbation. The helix can contain metastable irregularities as seen in the equilibrium state (c).
\label{Euler2helix}}
\end{center}
\end{figure}

To numerically study the solution path from the straight to the helical configuration, we introduce a tiny random perturbation to the straight compressed filament. A numerically obtained state diagram is depicted in Fig.~\ref{bifurcation-diagram}, where a suitable parameter distinguishing between the configurations is simply the projection of the filament length onto the $z$ axis. There exist two threshold forces beyond which the straight and the Euler-buckled states are unstable, respectively. The helical state is a consequence of the confinement and sets in after the filament has touched the wall. Above the threshold, the planar state remains a solution but is unstable with respect to the helix, which is demonstrated in Fig.~\ref{Euler2helix}. There one also sees a defect in the helix. Such defects are a general feature subject to the initial condition (perturbation) and are metastable, i.e., they are stable while the compressing force is not reduced.
\begin{figure}[h]
\begin{center}
\includegraphics[width=9cm]{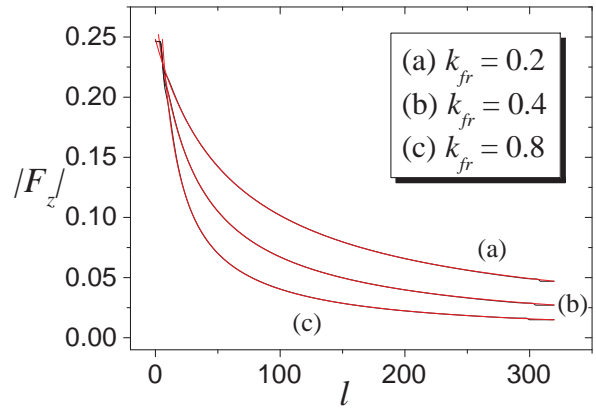}
\caption{(color online) Compression force profile $\vert F_z(l)\vert$ in a filament with length to radius ratio 320. The loading force is $\vert F_z(0)\vert=0.25 {E I/R^2}$, which is approximately 2,600-times the critical force for Euler buckling $F_{crit}= E I\pi^2/L^2$.
Numeric data (black) is excellently fit (red) with the function $f(l)=A (l_0+l)^c$ (the curves are perfectly overlapping on the scale of the figure): (a) $A=36.3$, $l_0=285.5$, $c=-0.882$, (b) $A=21.4$, $l_0=131.8$, $c=-0.894$, (c) $A=11.95$, $l_0=44.9$, $c=-0.903$.
\label{Fonl}}
\end{center}
\end{figure}
\begin{figure}[h]
\begin{center}
\vspace*{2cm}\includegraphics[width=8.5cm]{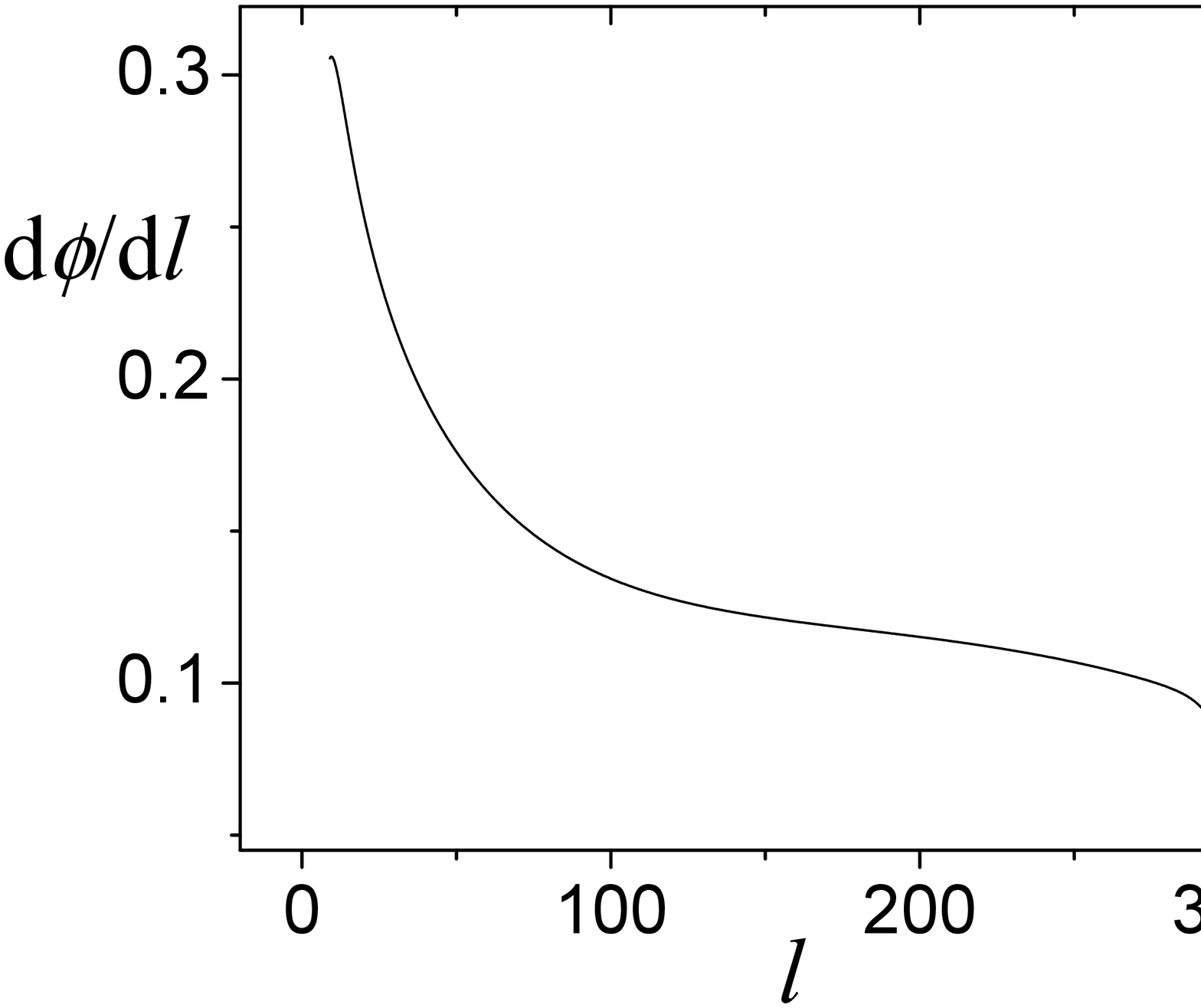}\vspace*{-2.5cm}
\caption{Profile of ${\rm d}\phi/{\rm d}l$, the inverse of the helical pitch $q_0 = 2\pi/({\rm d}\phi/{\rm d}l)$, for the filament of Fig.~\ref{Fonl} (c). The boundaries where the pitch is not defined are not displayed. As expected, $\dot{\phi}$ gets smaller (the pitch gets larger) as the longitudinal force is reduced. Its relative decrease is however smaller than that of the force, as hinted by the quadratic dependence $F_z\propto \dot{\phi}^2$ in Eq.~(\ref{dotphi}) of the simple model (which assumes constant pitch).
\label{pitch}}
\end{center}
\end{figure}

We are furthermore interested in the force that has to be applied to the lower end to keep it fixed, i.e., to sustain the confined helix. As we are to show, this force can be much less than the load applied to the upper end due to the action of friction and in fact approaches zero for very long confining channels. This opens up the possibility of jamming, i.e. of a stable static helical configuration of the rod that has been jammed against the confining walls via the surface friction and is sustained with only a tiny (zero in the limit of very long channel) opposing force at the other end of the channel.

\begin{figure}[h]
 \begin{center}
 \includegraphics[width=9cm]{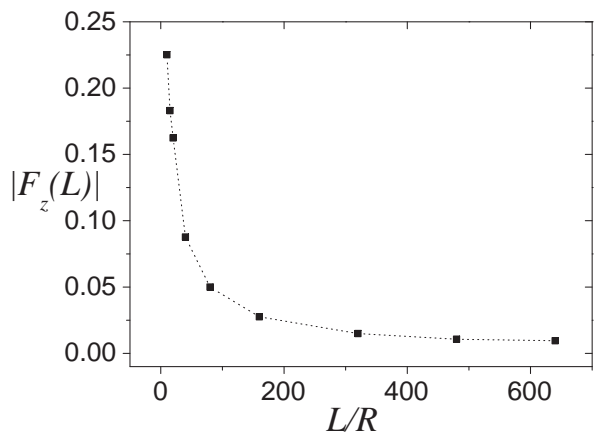}
 \caption{Compression force, $\vert F_z(L)\vert$, transmitted through filaments of lengths $L$; the external compressing force is $\vert F_z(0)\vert=0.25 {E I/R^2}$; $k_{fr}=0.8$. The dotted line serves as an eye guide.
 \label{FonL}}
\end{center}
\end{figure}

Fig.~\ref{Fonl} shows the profiles of the longitudinal force $\vert F_z\vert$ as we move along the filament, and Fig.~\ref{pitch} shows the profile of the helical deformation $\dot{\phi}$.
The force profiles are perfectly fit by a power law, yet with an exponent close to -0.9 instead of -1 as suggested by Eq.~(\ref{Fz-simple}).
The validity of the power law is remarkable though it deviates from the approximate universal scaling of Eq.~(\ref{Fz-universal}). In other words, the solution is practically indistinguishable from a power-law, in spite of the evidence that it considerably departs from our simple model. The reason for this remarkable validity of the power-law, yet with an exponent different than -1, is not clear. Fig.~\ref{FonL} shows the force transmitted to the lower end as the function of the length $L$ of the filament, at a fixed load on the upper end. We see that for long filaments the transmitted force is strongly reduced and asymptotically approaches zero.

\begin{figure}[h]
 \begin{center}
 \includegraphics[width=9cm]{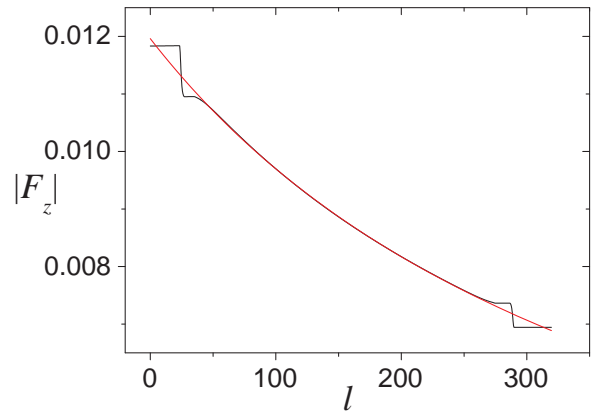}
 \caption{(color online) Compression force profile $\vert F_z(l)\vert$ (black) in the filament with length to radius ratio 320 for a smaller loading force $\vert F_z(0)\vert=0.0125 {E I/R^2}$ and $k_{fr}=0.8$, fit (red) with the function $f(l)=A (l_0+l)^c$; 
 $A=15.5$, $l_0=2010$, $c=-0.942$. 
 The reduction of the force is much smaller than for larger loads in Fig.~\ref{Fonl}, on account of the singular point $l_c$ in Eq.~(\ref{l_c}) moving further away. The exponent of the power law, however, remains nearly unchanged.
 The boundary regions get wider as the load is reduced.
\label{tiny_force}}
\end{center}
\end{figure}

\begin{figure}[h]
\begin{center}
	\mbox{\includegraphics[width=2.1cm]{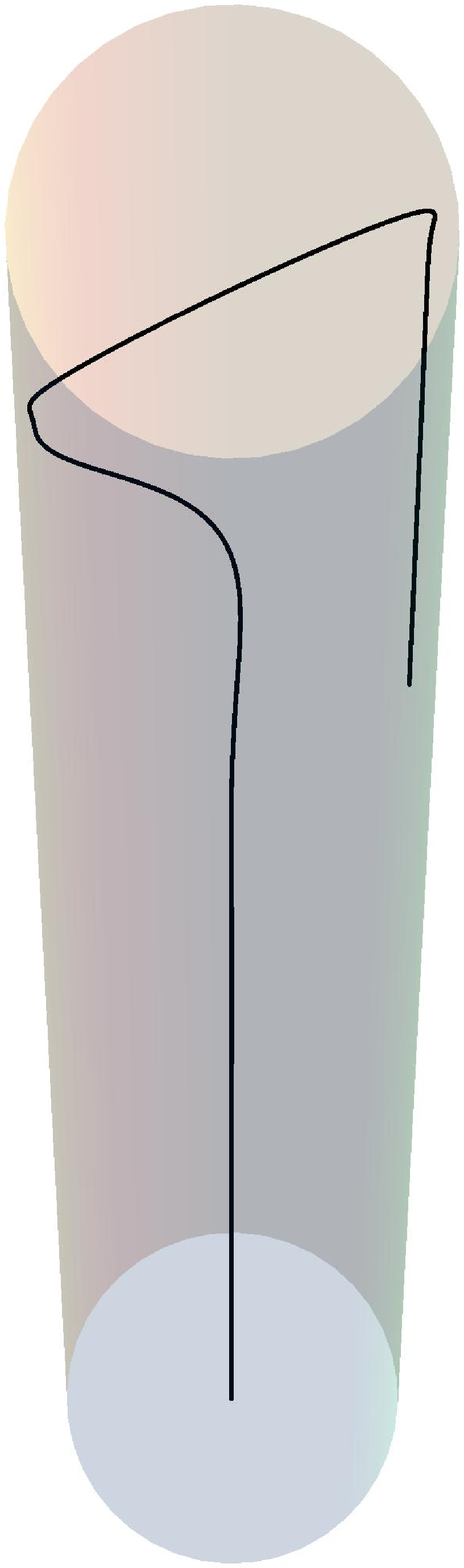}\hspace{0.5cm}
	      \includegraphics[width=2.1cm]{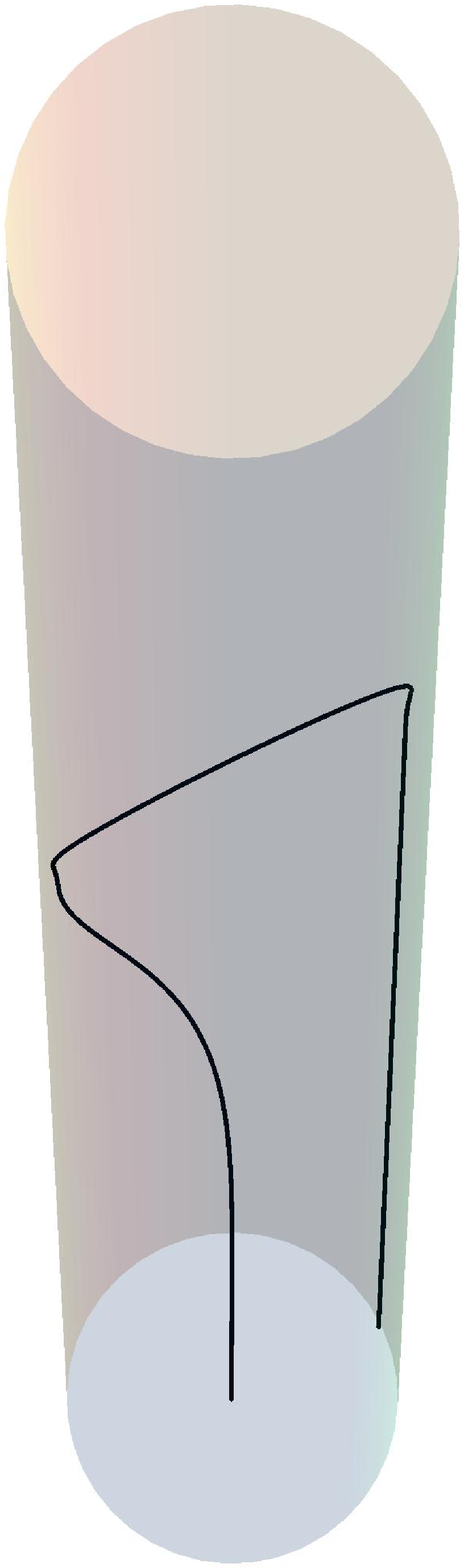}\hspace{0.5cm}
	      \includegraphics[width=2.1cm]{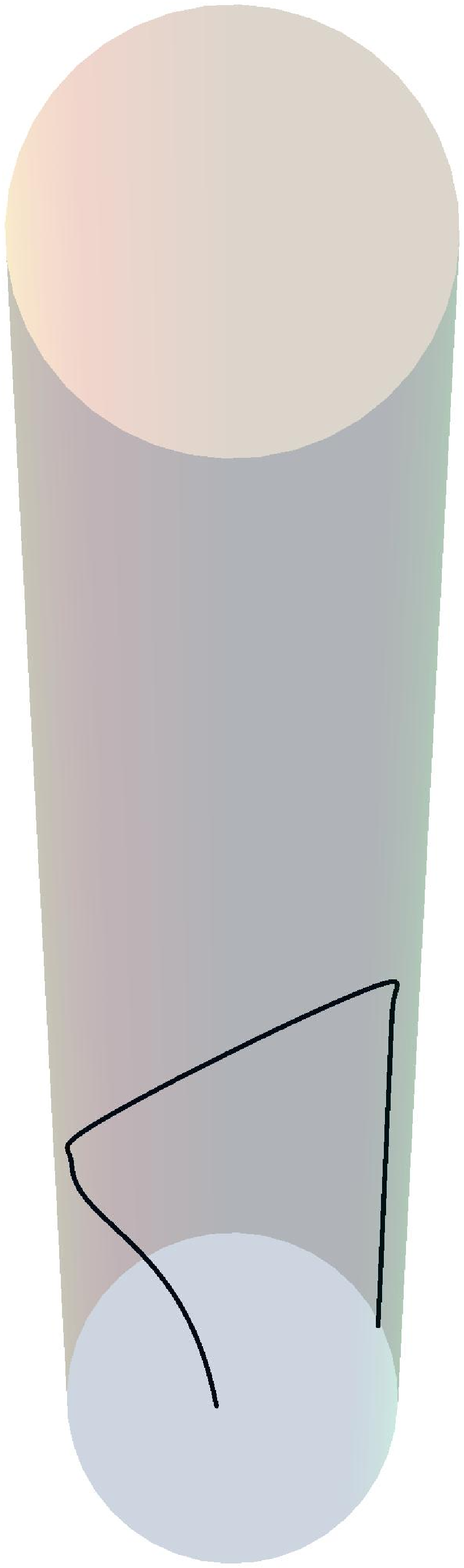}}
\caption{(color online) If the load is too large, the filament makes a U-bend and escapes out of the cylinder as shown on the subsequent figures instead of forming the helix. $L/R=320$, $\vert F_z(0)\vert=0.5 {E I/R^2}$, only the upper half of the cylinder is shown.
\label{large_force}}
\end{center}
\end{figure}

The physically interesting interval of loading force strengths depends on the aspect ratio of the filament. In Fig.~\ref{Fonl}, for example, the loading force was selected such that the transmitted force was significantly reduced. The reduction of the transmitted force is weaker if the loading force is smaller, Fig.~\ref{tiny_force}, in accord with Eq.~(\ref{l_c}) of the simple model. On the other hand, increasing the loading force beyond a threshold (that scales as $1/R^2$) results in a catastrophic event---a U bending of the filament, leading to the escape out of the confining pore, Fig.~\ref{large_force}. Hence, it is only for long and thin rods that the transmitted force can be strongly reduced.

\section{Discussion}

In this work we have analysed the behavior of an elastic filament that is confined in the longitudinal as well as transverse direction. The longitudinal confinement leads to the well known Euler buckling instability that depends on the external loading and the length of the filament. We have shown that the buckled configuration on touching the confining walls of a cylindrical channel then evolves through a planar deformed configuration towards a helical state via an additional instability. Considering the effects of the friction on the walls of the cylindrical enclosure we have been able to demonstrate, that the longitudinal force transmitted through the filament decays along its length. This phenomenon is not unrelated to the decaying longitudinal stresses within a cylindrical granular column anchored by the wall friction, or even a regular elastic solid with a finite Poisson ratio enclosed within a rigid hollow cylinder  and again anchored by wall friction. 

The important difference between the examples of granular and regular solids confined within cylindrical walls and the present case of thin elastic filament is the nature of the decay of the longitudinal force along the cylindrical enclosure. In the first two cases the decay is exponential and leads to a natural length scale for the problem. This means that the jamming of the granular or classical elastic bodies depends only on intrinsic parameters, describing the stress distributions within the body and the magnitude of the friction forces between the body and the cylindrical enclosure. In the case of the confined filament the longitudinal force transmitted along the cylindrical enclosure decays algebraically. In this case the onset of jamming is determined by the intrinsic elastic parameters of the filament as well as the longitudinal force counteracting the loading from the opposite side of the enclosure. This counter-force could be in principle very small and could even result from thermodynamic fluctuations on the other end of the enclosure if the cylindrical pore is small enough. Nevertheless the fact that the onset of jamming in the case of a confined elastic filament is scale free and thus depends on external constraints separates it fundamentally from the standard jamming in granular materials \cite{duran}.

What would be the systems that could exhibit this type of jamming scenario? 
We have found out that the criterion for the onset of helical postinstability shape is geometrical: according to the curve in Fig.~\ref{FonL}, the aspect ratio of polymer length (or persistence length, if it is smaller) and the radius of the pore should be of the order of 100 or more for a significant reduction of the compression force that could in principle lead to jamming. For carbon nanotubes with a typical length of $100\,\mu$m and Young's modulus of $10^{12}\,$Pa, the relevant pore radius is $1\,\mu$m or less, and the loading force is in the nN range. All reasonable values. On the other hand, DNA appears to be too flexible, i.e., its persistence length of 50\,nm is too short as it should be confined to a sub nanometer pore, violating the structural integrity if DNA (the diameter of DNA is approximately 2\,nm).  It thus appears that at least in principle one could observe the helical instability coupled to a jamming transition  within the context of confined microtubules. We hope that our theoretical work will provide enough motivation for experimentalists to search for this interesting phenomenon.

\section*{Acknowledgments}
This work has been supported by the Agency for Research and Development of Slovenia
under grants P1-0055(C), Z1-7171  and L2-7080.
Many thanks to Gregor Veble for fruitful discussions and useful hints. 

%\newpage


\begin{thebibliography}{99}

\bibitem{manning} G. S. Manning, Cell Biophys. {\bf 7}, 57-89 (1985).

\bibitem{podgornik-1} P. L. Hansen, D. Sven\v sek, V. A. Parsegian, R. Podgornik, Phys. Rev. E {\bf 60}, 1956-1966 (1999).

\bibitem{dogterom-1} M. Dogterom and B. Yurke, Science {\bf 278},  856-860 (1997).

\bibitem{lesne} J-M. Victor, E. Ben-Haim, and A. Lesne Phys. Rev. E {\bf 66}, 060901(R) (2002).

\bibitem{nanotube-inst} B. I. Yakobson, C. J. Brabec, and J. Bernholc, Phys. Rev. Lett. {\bf 76}, 2511 (1996).

\bibitem{nanorod-inst} E. W. Wong, P. E. Sheehan, and C. M. Lieber, Science {\bf 277}, 1971 (1997). 

\bibitem{dna-nanochannel} W. Reisner, K. J. Morton, R. Riehn, Yan Mei Wang, Zhaoning Yu, M. Rosen, J. C. Sturm, S. Y. Chou, E. Frey, and R. H. Austin, Phys. Rev. Lett. {\bf 94}, 196101 (2005). 

\bibitem{nanochan} F. Wagner, G. Lattanzi, and E. Frey, Phys. Rev. E {\bf 75}, 050902(R) (2007). 

\bibitem{odijk} T. Odijk, J. Chem. Phys. {\bf 125}, 204904 (2006).
 
\bibitem{janssen} H. A. Janssen, Z. Ver. Dt. Ing. {\bf 39}, 1045-1049 (1895).

\bibitem{elastic_validity} It can be shown that this is valid if the Poisson's ratio $\sigma$ is sufficiently small, $\sigma\ll 0.5$. 

\bibitem{landau} L. D. Landau and E. M. Lifshitz, {\it Theory of elasticity}, 3rd edition (Reed, Oxford, 1986).

\bibitem{lubinski} A. Lubinski, W. S. Althouse, J. L. Logan, J. Petrol. Tech. (June 1962), 655; Trans., AIME, 225.

\bibitem{mitchell1988}R. F. Mitchell, SPE Drilling Engineering (September 1988), 303.

\bibitem{mitchell1996}R. F. Mitchell, SPE Drilling \& Completion (September 1996), 178.

\bibitem{mitchell2002}R. F. Mitchell, SPE Journal (December 2002), 373.

\bibitem{love}A. E. Love, {\it A treatise on the mathematical theory of elasticity}, 4th edition (Dover Publications, New York, 1944).

\bibitem{duran} J. Duran, {\it Sands, Powders, and Grains: An Introduction to the Physics of Granular Materials}, 1st edition (Springer, 1999).

\end{thebibliography}
\end{document}